\documentclass[prd,twocolumn,showpacs,amsmath,amssymb,superscriptaddress,floatfix,nofootinbib]{revtex4}
\usepackage{amsfonts}
\usepackage[colorlinks=true, citecolor=blue, linkcolor=blue, urlcolor=blue]{hyperref}
\usepackage{color}
\usepackage{graphicx,amsfonts,multirow}
\usepackage{amsmath}
\usepackage{epstopdf,multirow}

\begin{document}

\title{Axion-like particle inflation and dark matter}

\begin{abstract}
In this paper, we propose a generalized natural inflation (GNI) model to study axion-like particle (ALP) inflation and dark matter (DM). GNI contains two additional parameters $(n_1, n_2)$ in comparison with the natural inflation, that make GNI more general. The $n_1$ build the connection between GNI and other ALP inflation model, $n_2$ controls the inflaton mass. After considering the cosmic microwave background and other cosmological observation limits, the model can realize small-field inflation with a wide mass range, and the ALP inflaton considering here can serve as the DM candidate for certain parameter spaces.
\end{abstract}

\author{Wei Cheng}\email{chengwei@itp.ac.cn}
\affiliation{School of Science, Chongqing University of Posts and Telecommunications, Chongqing 400065, P. R. China}
\affiliation{CAS key laboratory of theoretical Physics, Institute of Theoretical Physics, Chinese Academy of Sciences, Beijing 100190, China. University of Chinese Academy of Sciences, Beijing, 100190, China.}
\author{Ligong Bian}\email{lgbycl@cqu.edu.cn}
\affiliation{Department of Physics, Chongqing University, Chongqing 401331, China.}
\author{Yu-Feng Zhou}\email{yfzhou@itp.ac.cn}
\affiliation{CAS key laboratory of theoretical Physics, Institute of
Theoretical Physics, Chinese Academy of Sciences, Beijing 100190, China. University of Chinese Academy of Sciences, Beijing, 100190, China.}
\affiliation{School of Fundamental Physics and Mathematical Sciences,
Hangzhou Institute for Advanced Study, UCAS, Hangzhou 310024, China. }
\affiliation{International Centre for Theoretical Physics Asia-Pacific,
Beijing/Hangzhou, China.}


\maketitle

\section{Introduction}
In cosmology, inflation has been firmly established by the accurate cosmic microwave background (CMB), measurement about the temperature and polarization anisotropy~\cite{Ade:2015lrj,Story:2013,Hou:2014}, and the existence of dark matter (DM) also has been confirmed by numerous cosmological observations~\cite{Hinshaw:2012aka,Anton:1990,Sikivie:2006ni,Bertone:2004pz}, which lead a huge number of study to describe inflation and DM at the same time~\cite{Kofman:1994rk,Kofman:1997yn,Liddle:2006qz,Liddle:2008bm,Lerner:2009xg,DeSantiago:2011qb,Mukaida:2012bz,
Tenkanen:2016twd,Bastero-Gil:2015lga,Daido:2017wwb,Daido:2017tbr,Hooper:2018buz}. The important ingredient of connecting inflation and DM is a reheating process. Usually, the functions of inflaton are to drive the inflation and reheat the universe, while a new particle is added to act as the DM. To avoid adding new particles, this paper will extend the natural inflation (NI) to realize an axion-like particle (ALP) inflation, then the vast majority of inflationary energy is converted into radiation energy, and the remaining inflaton acts as DM, so as to achieve the ALP serving as both inflaton and DM while reheating the universe~\cite{Daido:2017wwb,Daido:2017tbr}.

The NI was first proposed in Ref.~\cite{Adams:1992bn} to solve the hierarchy problem of inflation. Its appendant property, shift symmetry, allows an ALP to act as the inflaton. Besides, it keeps the flatness of potential energy, which is essential for driving the expansion of the universe and generating density perturbations. The NI potential can be written as follows:
\begin{equation}
V(\phi)=\Lambda^4\bigg[1 + \cos(\frac{\phi}{f})\bigg],
\label{NI-model}
\end{equation}
where $\Lambda$ and $f$ are the inflaton energy density and decay constant. However, the model requires a Super-Planckian decay constant, which may not be justified in the framework of quantum gravity, while the feasible parameter space of this simple NI is still shrinking according to the latest CMB data~\cite{Akrami:2018odb}. Thus, it has inspired many extension versions for the NI potential, such as: the Multi-NI model~\cite{Czerny:2014wza,Takahashi:2019qmh} and other generalization of natural inflation model~\cite{Munoz:2014eqa}, etc.

As a new attempt, we construct a generalized natural inflation (GNI) model to realize a single ALP natural inflation. In GNI, the new parameter $\epsilon$ is proportional to the decay constant $f$, thus larger (smaller) $\epsilon$ may lead to large (small)-field inflation. It is also proportional to the ALP mass, which will enable the mass to go from Sub-eV to Super-GeV. Moreover, GNI can be reduced to the previous models by fixing some parameters, which are elaborated in the main body.

After the inflation finished, the universe was stretched and diluted extremely, thus the temperature of the universe became very cool so that a reheating process is required to reheat the universe~\cite{Albrecht:1982mp,Linde:1982bu}. Many models devote to describe this process, such as perturbation reheating~\cite{Abbott:1982hn,Dolgov:1982th,Dolgov:1982qq,Cook:2015vqa}, non-perturbation reheating~\cite{Allahverdi:2011aj,Koshelev:2017tvv}, and instant reheating~\cite{Felder:1998vq,Kurkela:2011ti}. In this paper, the reheating of the universe is implemented by ALP inflaton decaying into Standard Model (SM) photon particles. It is worth noting that for large mass ALP inflaton, it will decay quickly and cannot serve as the DM, but for small mass, ALP inflaton can be a long-lived DM, but the universe cannot be directly reheated using the usual methods. In order to solve this problem, we find that the inflation field is a damped oscillation motion near the minimal potential energy according to its equation of motion, that is, the inflationary field begins to oscillate at a certain time after the inflation finished, and this oscillation is the damped oscillation of amplitude attenuation. If the mass in the ALP decay rate is an effective mass that is related to the oscillation amplitude~\cite{Daido:2017wwb,Daido:2017tbr}. Thus, the decay rate at the beginning of the oscillation may be very large, and then decreases rapidly, which not only solves the problem of reheating the universe with ALP inflaton but also makes it possible for ALP inflaton to act as long-lived DM at the same time~
\cite{Daido:2017wwb,Daido:2017tbr,Kofman:1994rk,Kofman:1997yn,Cardenas:2007xh,Hooper:2018buz,Choi:2019osi}.

This work is organized as follows: in Sec.~\ref{Model}, GNI is described along extend the NI model,
In Sec.~\ref{Inflation}, we show the ingredients of the slow-roll inflation and the inflaton mass analytically in GNI, and we discuss the source of the massive depression effect.
In Sec.~\ref{DM}, we calculate the evolution of energy density in the reheating phase and further discuss the ALP-DM relic density.
Finally, we briefly summary in Sec.~\ref{Summary}.

\section{The model}\label{Model}

In this section, we will elaborate on GNI and its property.

\subsection{Inflationary potential}

We extend the NI potential Eq.~(\ref{NI-model}) as follows:
\begin{equation}
V(\phi)=\Lambda^4\bigg[\cos(\frac{\phi}{f})+\epsilon \cos(\frac{\phi}{f})+\exp{[\frac{1}{n_1!}\cos(\frac{\phi}{f}+\frac{\pi}{n_1!})^{n_1}]}\bigg]^{n_2}+C\;,\label{eq:pot}
\end{equation}
where the $C$ is a constant that shifts the minimum of the potential to zero, the term $\epsilon \cos(\frac{\phi}{f})$ is the extension for the $\cos(\frac{\phi}{f})$ term with the value of $\epsilon$ closing to $0$. In addition, the positiveness of inflaton mass lead to $\epsilon >0$. The exponential term is a slightly change for the number $1$ in Eq.~(\ref{NI-model}). And the positive integers $n_1$ and $n_2$ make GNI more universal. As will be shown latter, $n_2$ controls the ALP mass and is highly related with DM property. When we take $n_1=0$ and $n_2=1$, it will reduce to the NI model. When we take $n_1=0$, it will reduce to other generalization of natural inflation model~\cite{Munoz:2014eqa}. Moreover, the parameter $n_1$ can be fixed as $1$ or $2$ considering the following analysis.

GNI is a slight extension of the NI model, $\phi=0$ and $\phi=\pi f$ are the maximum and minimum points for the GNI potential, respectively. Which yields:
\begin{align}
V'(\phi)|_{\phi=0}&=0,~~~V'(\phi)|_{\phi=\pi f}=0,\label{con1}\\
V''(\phi)|_{\phi=0}&<0,~~~V''(\phi)|_{\phi=\pi f}>0,\label{con2}
\end{align}
where the prime denotes the derivative with respect to $\phi$. With the potential being given by Eq.~(\ref{eq:pot}), we have
\begin{widetext}
\begin{align}
&V'(\phi)=\frac{n_2{\Lambda ^4}{{\bigg[{{e}^{\frac{{\cos{{(\frac{\phi }{f} + \frac{\pi }{{n_1!}})}^{n_1}}}}{{n_1!}}}} + (1+\epsilon )\cos(\frac{\phi }{f})\bigg]}^{-1 + n_2}}\bigg[(1+\epsilon ) n_1!\sin(\frac{\phi }{f}) + {{\rm{e}}^{\frac{{\cos{{(\frac{\phi }{f} + \frac{\pi }{{n_1!}})}^{n_1}}}}{{n_1!}}}}n_1\cos{{(\frac{\phi }{f} + \frac{\pi }{{n_1!}})}^{-1 + n_1}}\sin(\frac{\phi }{f} + \frac{\pi }{{n_1!}})\bigg]}{fn_1!},\\
\label{po:oneD}
&V'(\phi)|_{\phi=0}=- \frac{{{e^{\frac{{\cos{{(\frac{\pi }{{n_1!}})}^{n_1}}}}{{n_1!}}}}{{\bigg(1+\epsilon + \exp^{\frac{{\cos{{(\frac{\pi }{n_1!})}^{n_1}}}}{n_1!}}\bigg)}^{ - 1 + n_2}}n_1n_2{\Lambda ^4}\cos{{(\frac{\pi }{{n_1!}})}^{ - 1 + n_1}}\sin(\frac{\pi }{{n_1!}})}}{{fn_1!}},\\
&V'(\phi)|_{\phi=\pi f}=-\frac{{{e^{\frac{{-\cos{{(\frac{\pi }{{n_1!}})}^{n_1}}}}{{n_1!}}}}{{\bigg(-1-\epsilon + {{\exp}^{\frac{{-\cos(\frac{\pi }{{n_1!}})}^{n_1}}{{n_1!}}}}\bigg)}^{ - 1 + n_2}}n_1n_2{\Lambda ^4}(-\cos{{(\frac{\pi }{{n_1!}})})^{ n_1}}\tan(\frac{\pi }{{n_1!}})}}{{f n_1!}}.
\end{align}
\end{widetext}
The first condition in Eq.~(\ref{con1}) leads to $n_1=0, 1$ and $2$.
For $n_1=0$, GNI will reduce to other generalization of natural inflation model~\cite{Munoz:2014eqa}, and we will not discuss it.
The $n_1=1$ scenario can not fulfill the second condition in Eq.~(\ref{con2}), and the last scenario $n_1=2$ satisfy all of the conditions given in Eq.~(\ref{con2}).

\begin{figure*}[!htp]
\begin{center}
\includegraphics[scale=0.482]{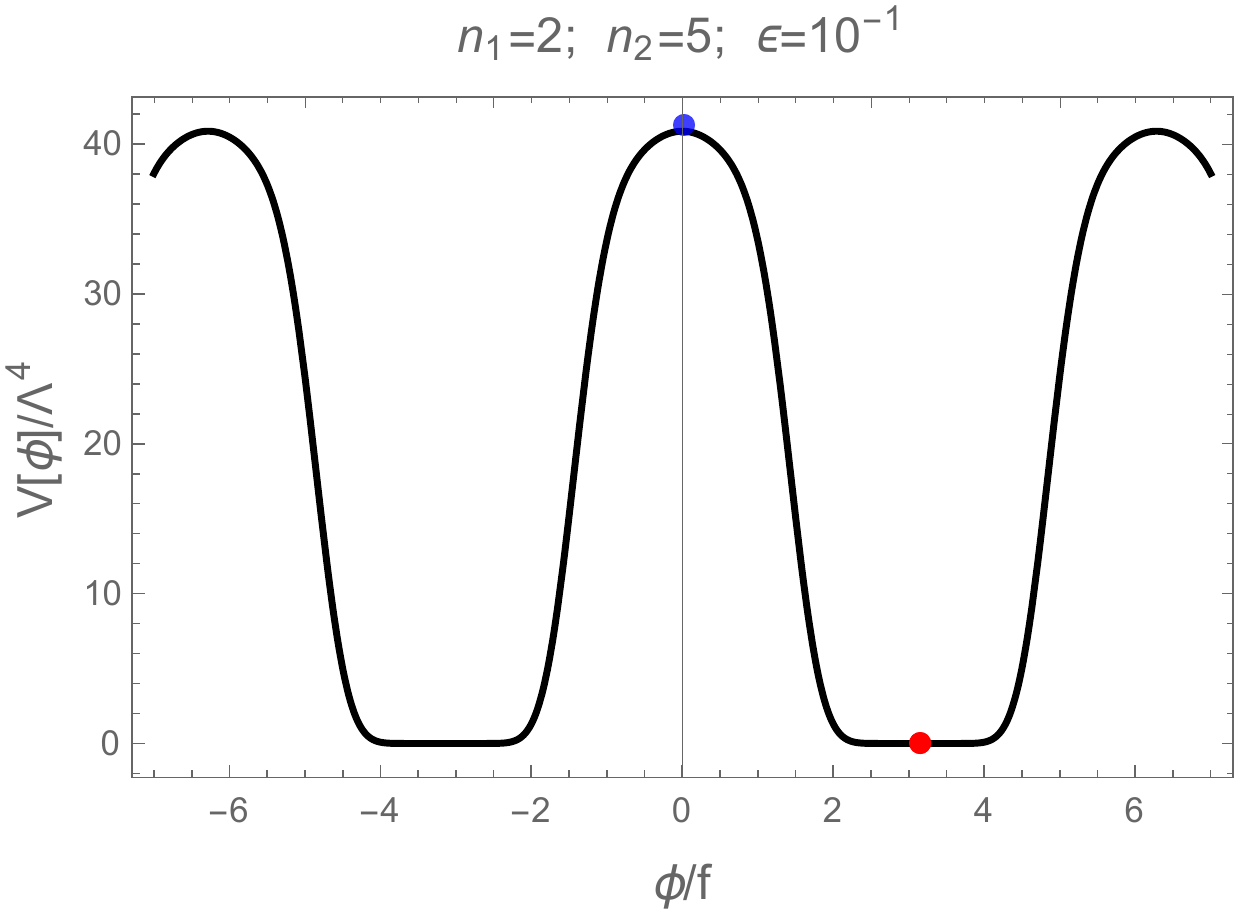}
\includegraphics[scale=0.54]{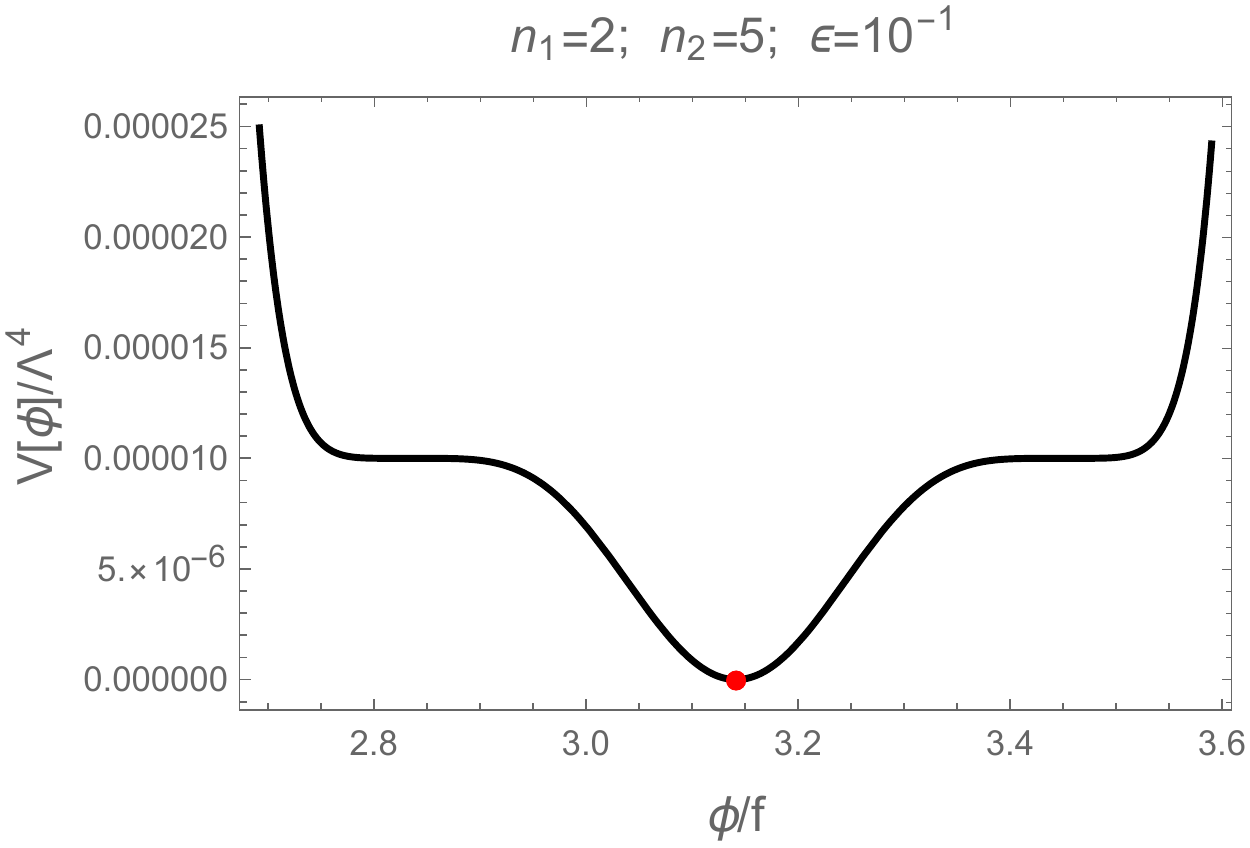}
\includegraphics[scale=0.482]{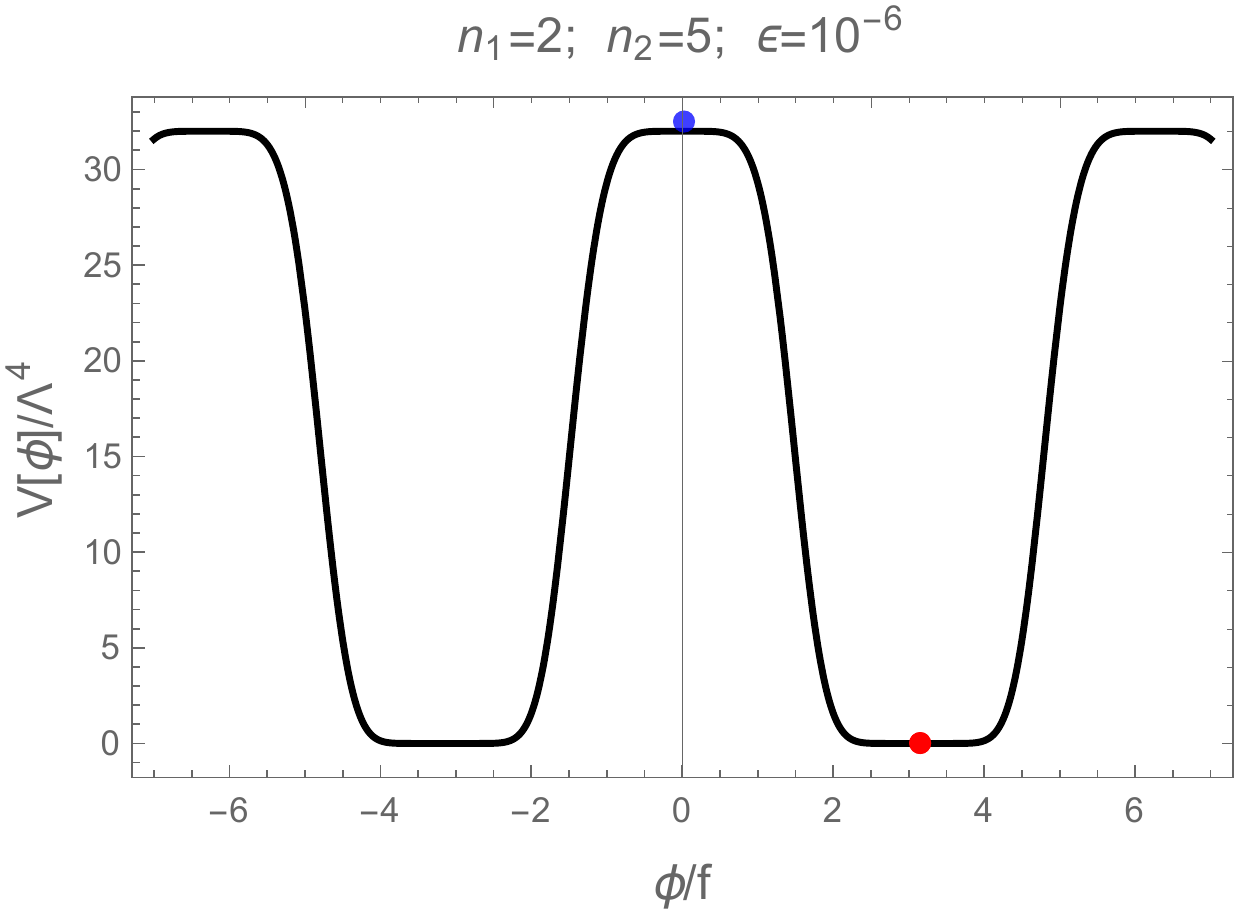}
\includegraphics[scale=0.54]{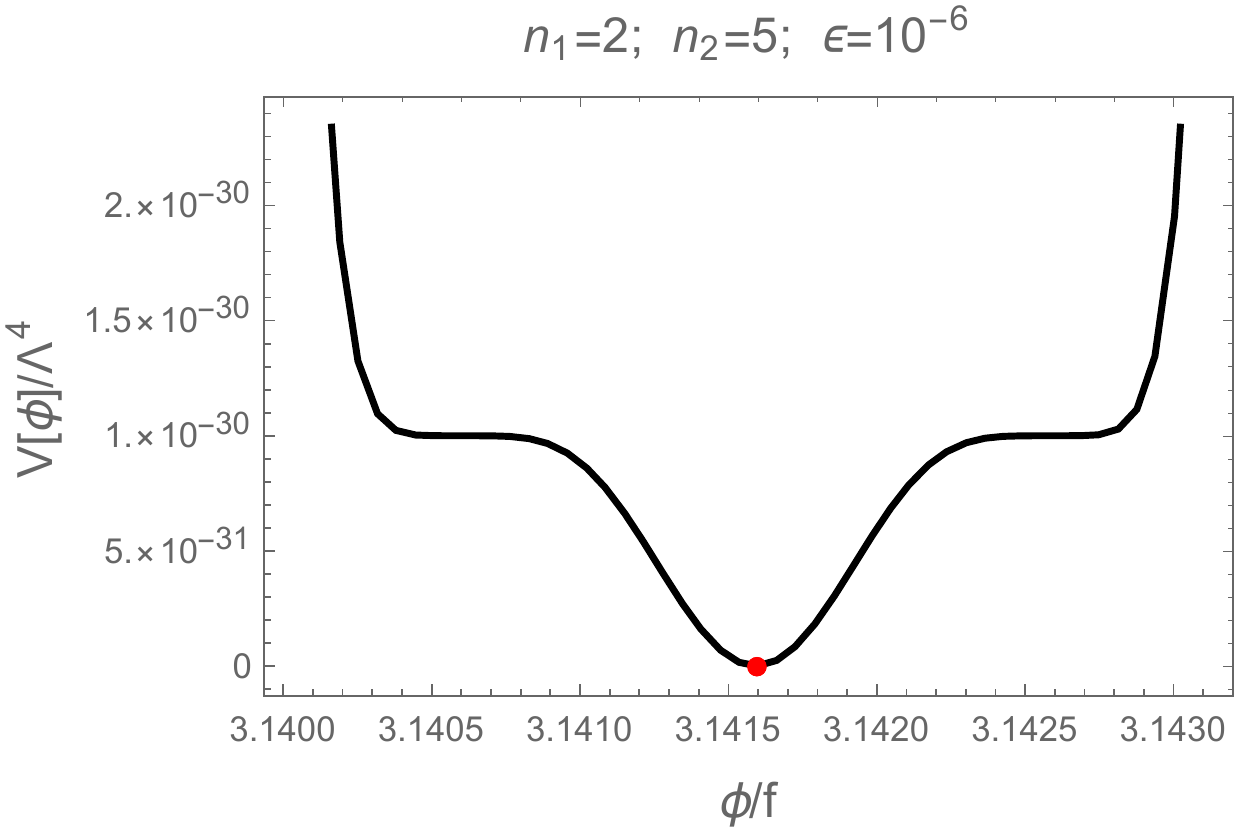}
\caption {Potential shape for extend Natural-Inflation. Left: Small (Large) field inflation potential for the bottom (top) panel. Right: Corresponding true vacuum points. The red points represent the true vacuum points, and slow-roll inflation is possible around the blue dot.
}
\label{fig:LS-inf}
\end{center}
\end{figure*}

For $n_1=2$, the second derivative of the potential can be written:
\begin{equation}
V''(\phi)|_{\phi=0}=- \frac{{\epsilon {{(2+\epsilon)}^{ - 1 + n_2}}n_2{\Lambda ^4}}}{{{f^2}}},
\end{equation}
\begin{equation}
V''(\phi)|_{\phi=\pi f}=\frac{{{{(-\epsilon)}^{ - 1 + n_2}}(2+\epsilon)n_2{\Lambda ^4}}}{{{f^2}}}\;.
\end{equation}
As $V''(\phi)|_{\phi=0}<0$ and $V''(\phi)|_{\phi=\pi f}>0$, so $\epsilon>0$ and $n_2$ is odd number.
Furthermore, two inflection points around the $\phi=\pi f$ points can be obtained in terms of Eq.~(\ref{po:oneD}). 
This also can be seen from the right panels of Fig.~\ref{fig:LS-inf}.

\begin{figure*}[thb]
\begin{center}
\includegraphics[scale=0.55]{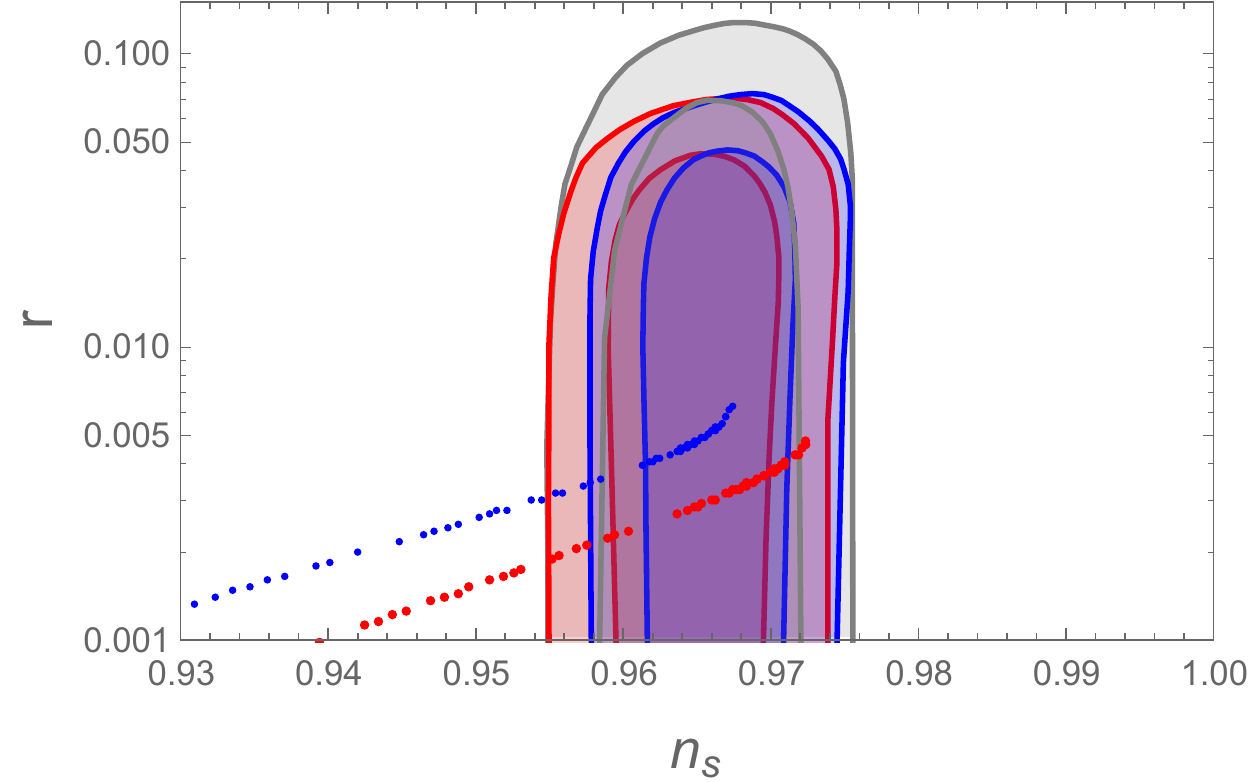}
\includegraphics[scale=0.55]{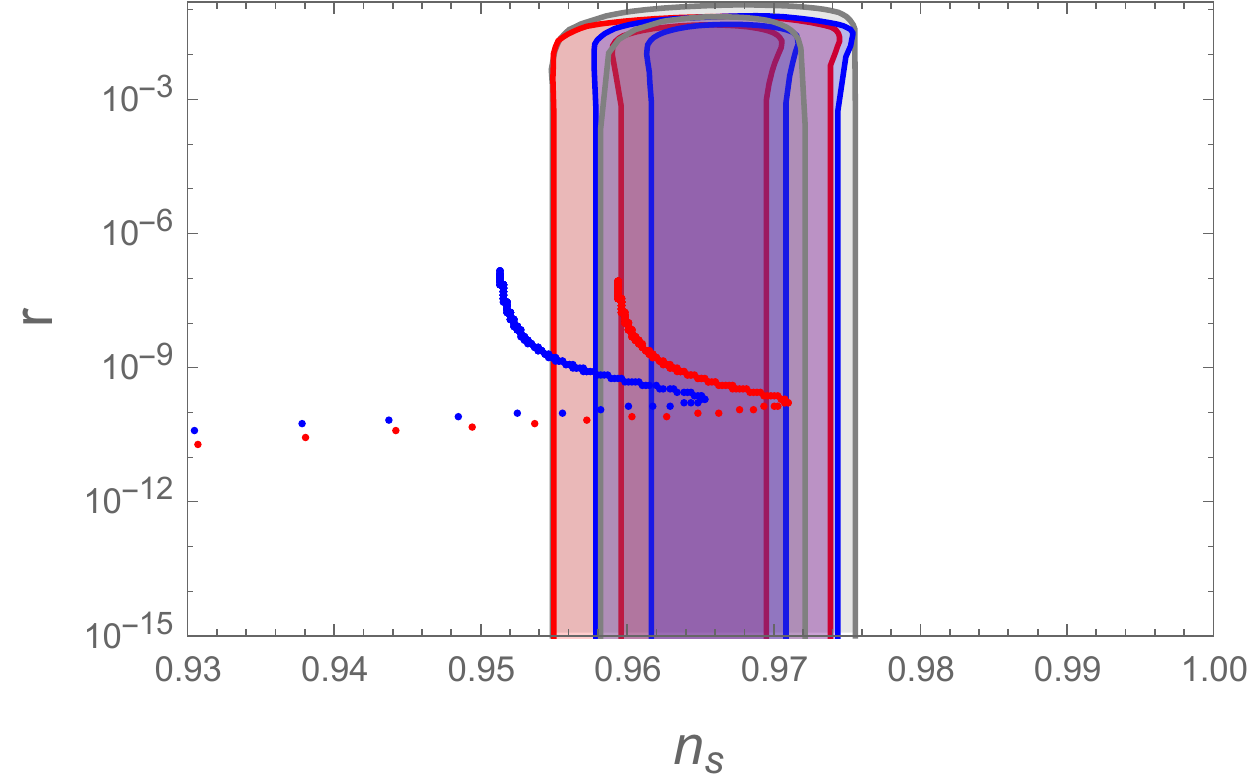}
\caption {Left: Large-field inflation predictions in $(n_s,r)$ planes with $\epsilon=10^{-1}$, $n_2=5$.
Right: Small-field inflation predictions in $(n_s,r)$ planes with $\epsilon=10^{-6}$, $n_2=5$.
Red (blue) points correspond to the e-folding number $N=60$ ($N=50$).
Gray, red and blue shadows stand for [TT,TE,EE+lowE+lensing], [TT,TE,EE+lowE+lensing+BK15] and [TT,TE,EE+lowE+lensing+BK15+BAO] constraints~ \cite{Akrami:2018odb}.}
\label{fig:nsr1}
\end{center}
\end{figure*}

The large field inflation will destroy the flatness of the potential and generate gravitational wave~\cite{Linde:1983gd}, while the small field inflation is reversed. Note that both large- and small-field inflation can be achieved in GNI, which are shown obviously in the left panels of Fig.~\ref{fig:LS-inf}. For the study of tensor-to-scalar ratio $r$ for large- and small-field inflation, we left it in the next subsection.

\subsection{ALP inflation}\label{Inflation}
We briefly review the slow-roll inflation in this section.
Slow-roll inflation has two characteristics: the kinetic energy of inflaton is much less than that of the potential energy, and the change rate of kinetic energy is much less than that of the cosmic expansion. The two slow-roll parameters are
\begin{equation}
\varepsilon(\phi) \equiv \frac{{M_{pl}}^2}{2}\left(\frac{V'(\phi)}{V(\phi)}\right)^2 ,\\
\quad \eta(\phi) \equiv {M_{pl}}^2\left(\frac{V''(\phi)}{V(\phi)}\right),
\end{equation}
where the prime denotes the derivative with respect to $\phi$.
During the inflation, $\varepsilon(\phi) \ll 1,$ and $\eta(\phi) \ll 1$. If anyone of the conditions breaks, i.e. $\varepsilon=1$ or $\mid\eta\mid=1$, the inflation will be finished. This can be used to set the field value ($\phi_{end}$) at the end of inflation.

The e-folding number $N_{I}$ is written as:
\begin{align}\label{eqq1}
N_I = \int H dt.
\end{align}
This will give the initial value $\phi_{I}$ for a fixed e-folding number. We will take $N_I=60$ to conduct our calculations.

Up to the first order level, the scalar spectral index $n_s$ and the tensor-to-scalar ratio $r$ can be expressed as:
\begin{equation}
\label{ns}
  n_s = 1-6\varepsilon+2\eta
\end{equation}
and
\begin{equation}
  r = 16\varepsilon.
\end{equation}
Here $n_s$ is dominated by the shape of the inflaton potential, $r$ is associated with inflationary energy scale~\cite{Czerny:2014wza}.

The tensor-to-scalar ratio $r$ as well as the scalar spectral index $n_s$ are tightly constrained by the Planck data combined with CMB observations as~\cite{Akrami:2018odb}:
\begin{align}
n_s &= 0.9649 \pm 0.0042, \\
r &< 0.10 ~~~(95\% {\rm CL}).
\end{align}

There is another important experimental observation, i.e., amplitude of scalar fluctuations $\Delta_\mathcal{R}^2$,
\begin{equation}
\Delta_\mathcal{R}^2 = \frac{1}{24 \pi^2 M_{\rm pl}^4}\frac{V(\phi)}{\varepsilon(\phi)}.
\end{equation}
where $\Delta_{\mathcal R} \simeq e^{3.098} \times 10^{-10} \simeq 2.2 \times 10^{-9}$~\cite{Ade:2015lrj}, which will limit the height of the potential, and may eliminate the suppressed effect from the pre-parameter of the mass for the usual extended NI, this is the reason that there is a large mass for the usual extended NI.

For illustration, in Fig.~\ref{fig:nsr1}, we show that both large- and small-field inflation can be realized for $n_2=5$. The small $\epsilon$ corresponds to small decay constant $f$, and then leads to small-field inflation. The large-field inflation may generate sizable tensor perturbations, which further lead to sizable gravity waves that are beyond the scope of this work.


\begin{figure*}[thb]
\begin{center}
\includegraphics[scale=0.33]{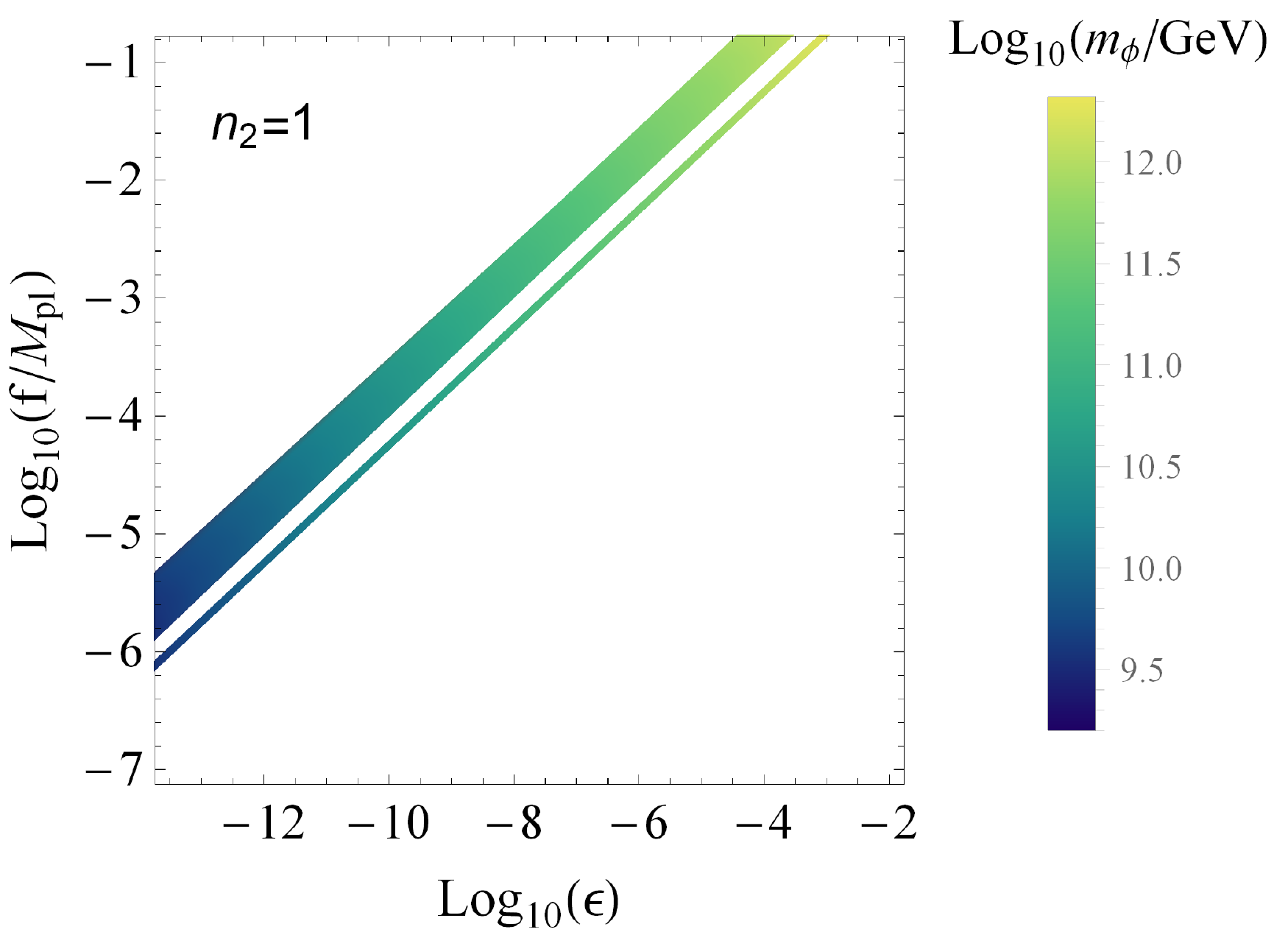}
\includegraphics[scale=0.33]{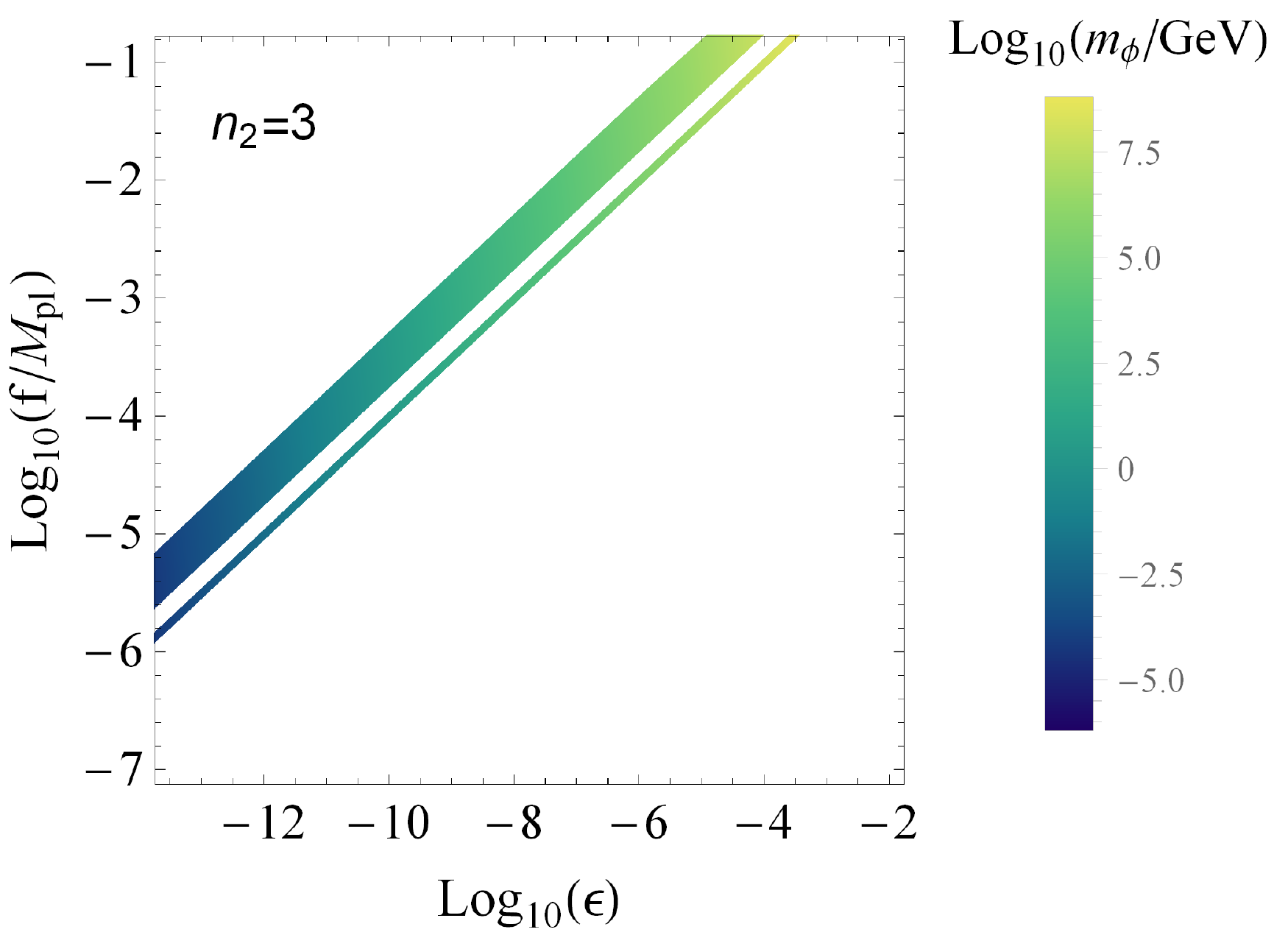}
\includegraphics[scale=0.33]{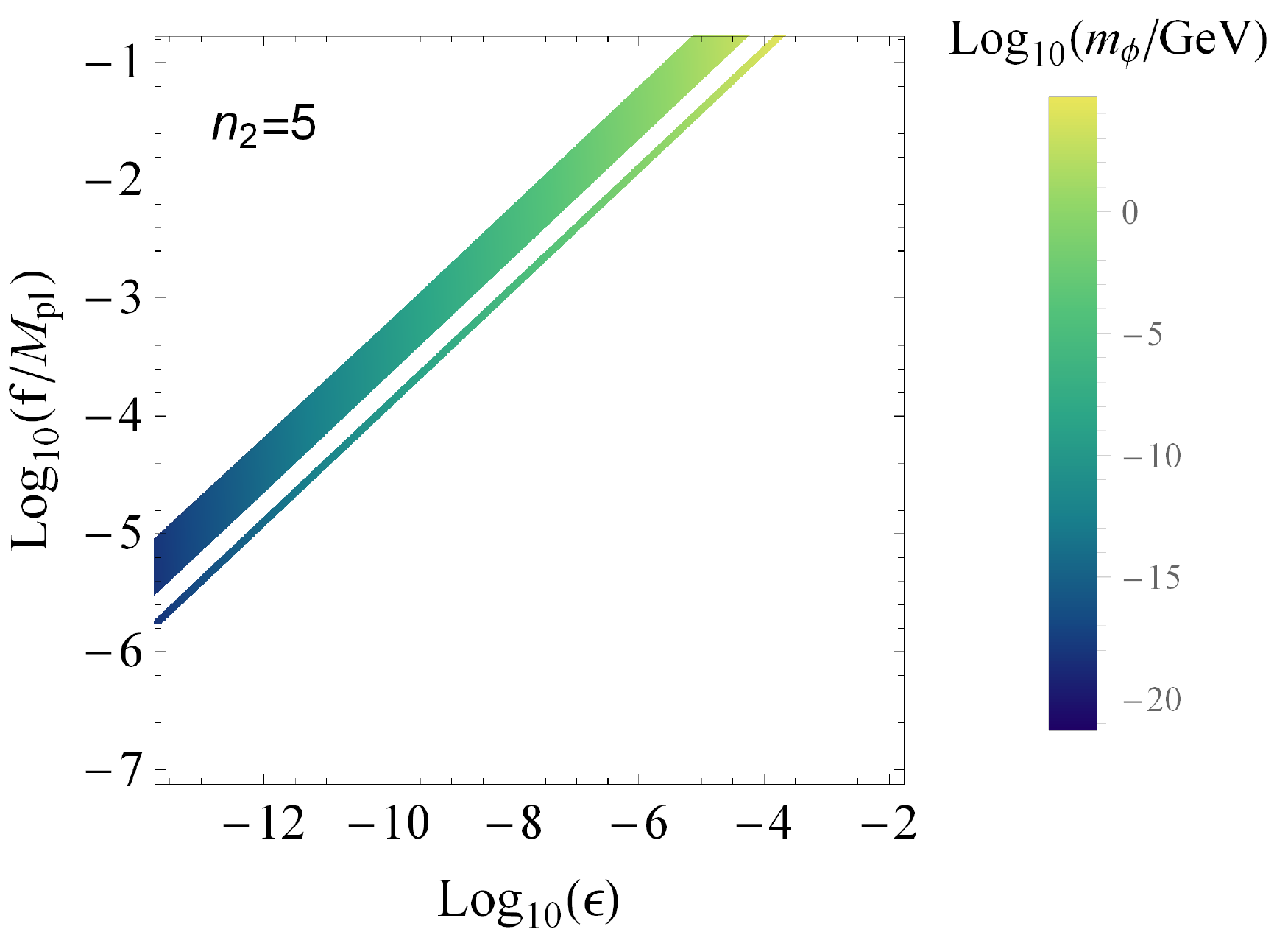}
\includegraphics[scale=0.33]{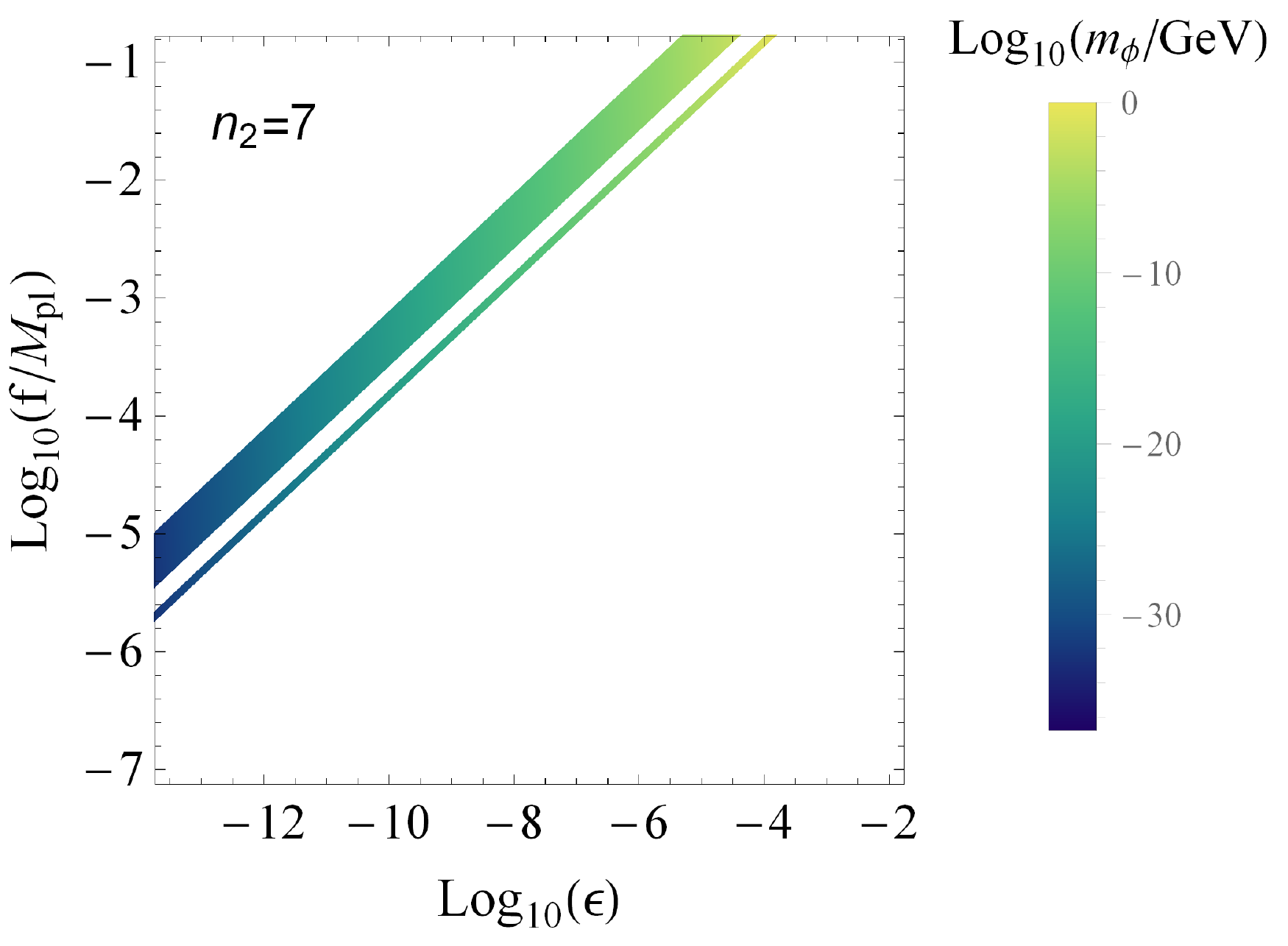}
\includegraphics[scale=0.33]{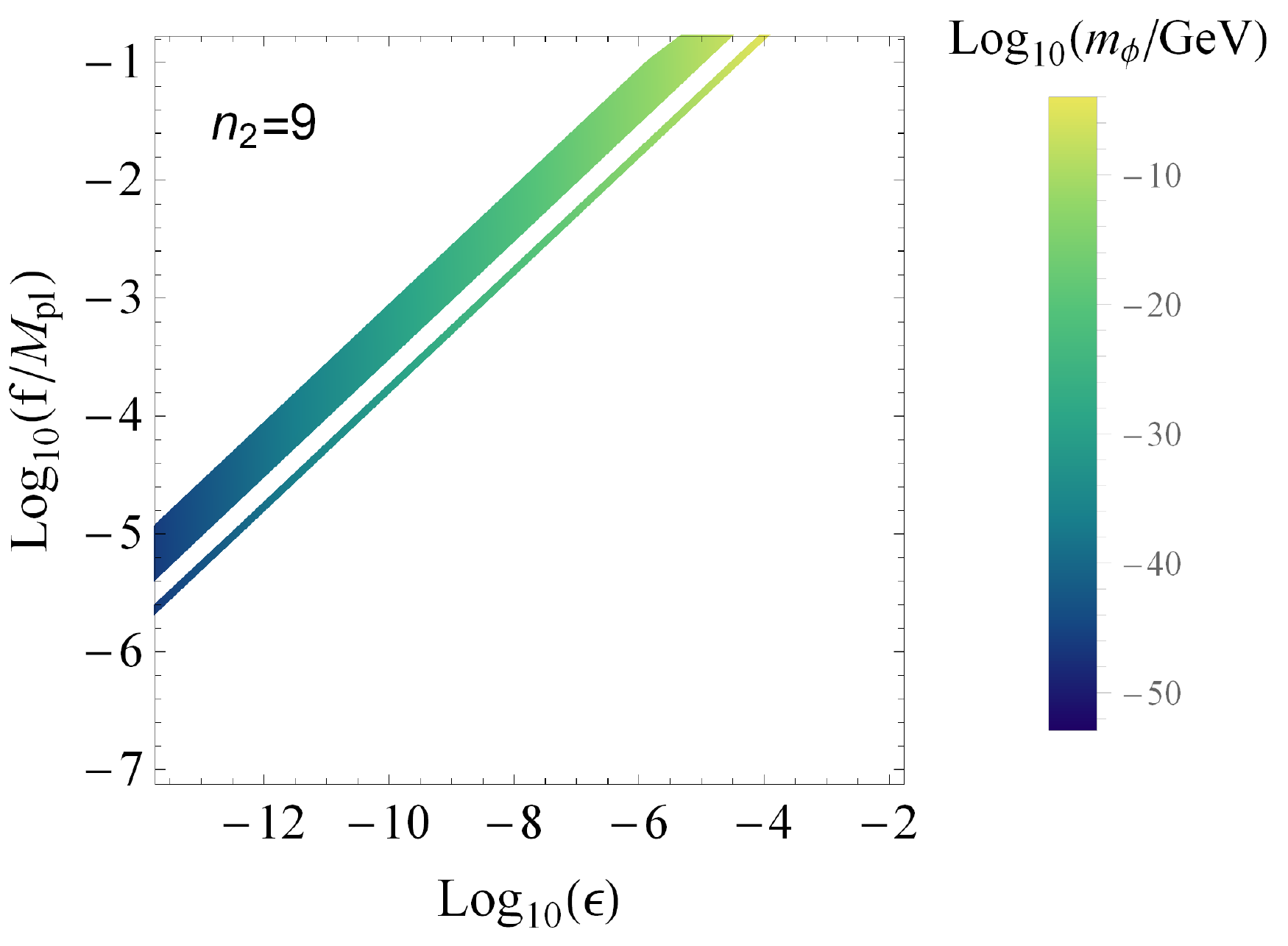}
\caption {The decay constant $f$ as a function of parameter $\epsilon$ for several values of $n_2$ with $n_1=2$. All the shades are the feasible regions under the latest CMB data constraints. Note that the boundary of those shades are caused by the limit of $n_s = 0.9649 \pm 0.0042$.}
\label{fig:nsr2}
\end{center}
\end{figure*}

To investigate the inflaton mass, we need to study the second derivative of potential at the true vacuum, i.e., minimal point of potential $(\phi=\pi f)$. The inflaton mass can be written as:
\begin{equation}
m_\phi=n_2^{1/2}(2+\epsilon)^{1/2}(-\epsilon)^{(n_2-1)/2}\frac{\Lambda^2}{f}\;.
\label{mass}
\end{equation}
 This indicates that the ALP mass is discrete for different odd integer $n_2$. In addition, the ALP mass will decrease rapidly with the increase of $n_2$ due to the factor $(-\epsilon)^{(n_2-1)/2}$, which makes a wide range for ALP mass.

We plot the ALP mass as a function of the decay constant $f$ and $\epsilon$ for several values of $n_2$ with $n_1=2$ in Fig.~\ref{fig:nsr2}, in which all shadows are confined by CMB data. For a fixed $n_2$, the decay constant $f$ increases monotonically with the parameter $\epsilon$, and the narrow space between two shadow bands are caused by the limit of $n_s = 0.9649 \pm 0.0042$~\cite{Akrami:2018odb}. With the increasing of $n_2$, the decay constant $f$ and the parameter $\epsilon$ remains almost unchanged, while the mass falling fast and extremely small inflaton mass can also be obtained. The reason is that there is a suppression effect for the mass from the power exponent ($n_2$), i.e., mass $\sim(-\epsilon)^{(n_2-1)/2}$ with $0<\epsilon<1$ in term of Eq.~(\ref{mass}).

\section{Dark matter}\label{DM}

In this paper, we only consider ALP inflaton coupled to photon, the Lagrangian can be written as~\cite{Blinov:2019rhb}:
\begin{equation}
\mathcal{L}_{\rm{ALP}}=\frac{1}{2}\partial_\mu \phi \partial^\mu \phi - \frac{1}{2}m^2_\phi \phi^2-\frac{1}{4}g_{\phi \gamma\gamma} \phi F_{\mu\nu}\tilde{F}^{\mu\nu},
\end{equation}
where $\tilde{F}^{\mu\nu}$ is the dual electromagnetic field-strength tensor, and the coupling $g_{\phi \gamma\gamma}=c_r\alpha/(2\pi f)$ with the ALP decay constant $f$. We fix $c_r$ in the range of $(10^{-3} - 10^{2})$  to do our numerical analysis.
The perturbative decay rate can be obtained as follows:
\begin{equation}
\Gamma_0 (\phi \to \gamma \gamma) = \frac{\alpha^2 c_\gamma^2}{64 \pi^3} \frac{m_\phi^3}{f^2}.
\end{equation}

Reheating epoch is a process that follows inflation, and the motion of the field in this epoch is actually a damped oscillation. Specifically, the equation of motion for the ALP zero mode in the early universe is
\begin{align}
\ddot{\phi} + 3H \dot{\phi} + m_\phi^2 \phi = 0,
\end{align}
where the Hubble parameter $H^2 =(8\pi{\rho_{tot}}/{3 M_{p}^2})$. When the Hubble parameter becomes comparable to $m_\phi$, i.e., $m_\phi=qH$ with $q\simeq 3$, the field begins to oscillate with initial energy density $\rho_\phi= \frac{1}{2}m_\phi^2 f^2\theta_0^2$, which will affect the DM relic abundance today~\cite{Blinov:2019rhb}{\footnote{Here $\theta_0$ is the initial misalignment angle, usually $\theta_0 \sim $$\cal{O}$ $(1)$.}}. The solution for the above field equation of motion can be written as the following oscillation form~\cite{Cardenas:2007xh}:
\begin{align}
\phi(t)=\frac{\Phi_0 M_{pl}}{m_{\phi} t}\sin(m_{\phi} t).
\label{Eq:fieldosc}
\end{align}

We can obtain the effective ALP decay width in the reheating epoch as follows:
assuming the ALP mass is an effective mass that is related to the amplitude of the inflationary potential, i.e., $m_{eff}^2=V''(\phi_{amp})$, and replacing the mass $m_\phi$ in the decay rate $\Gamma_0 (\phi \to \gamma \gamma)$ with $m_{eff}$, the new decay rate becomes $\Gamma_{eff} (\phi \to \gamma \gamma)$.
\begin{figure*}[t]
\begin{center}
\includegraphics[scale=0.35]{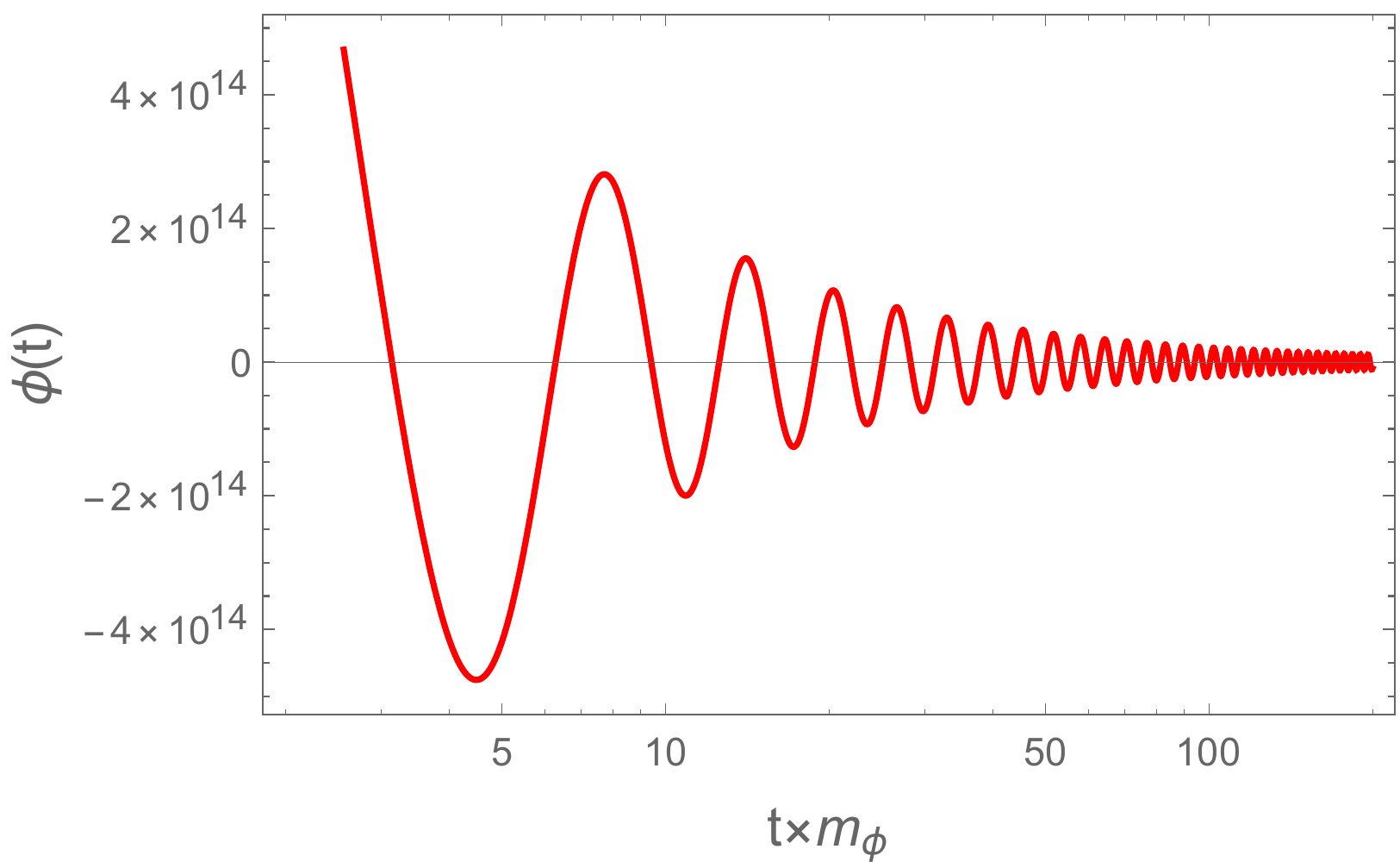}
\includegraphics[scale=0.36]{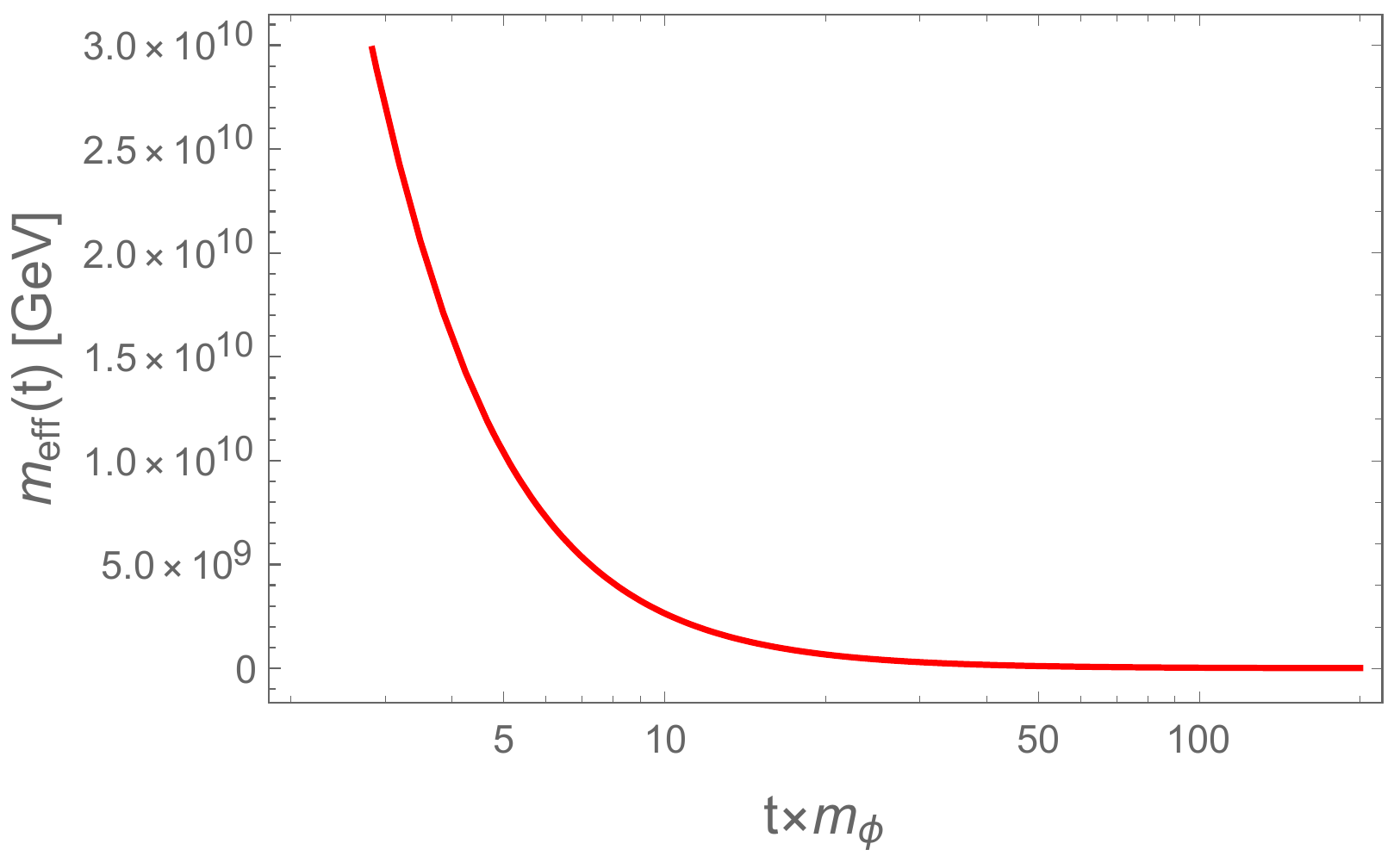}
\includegraphics[scale=0.36]{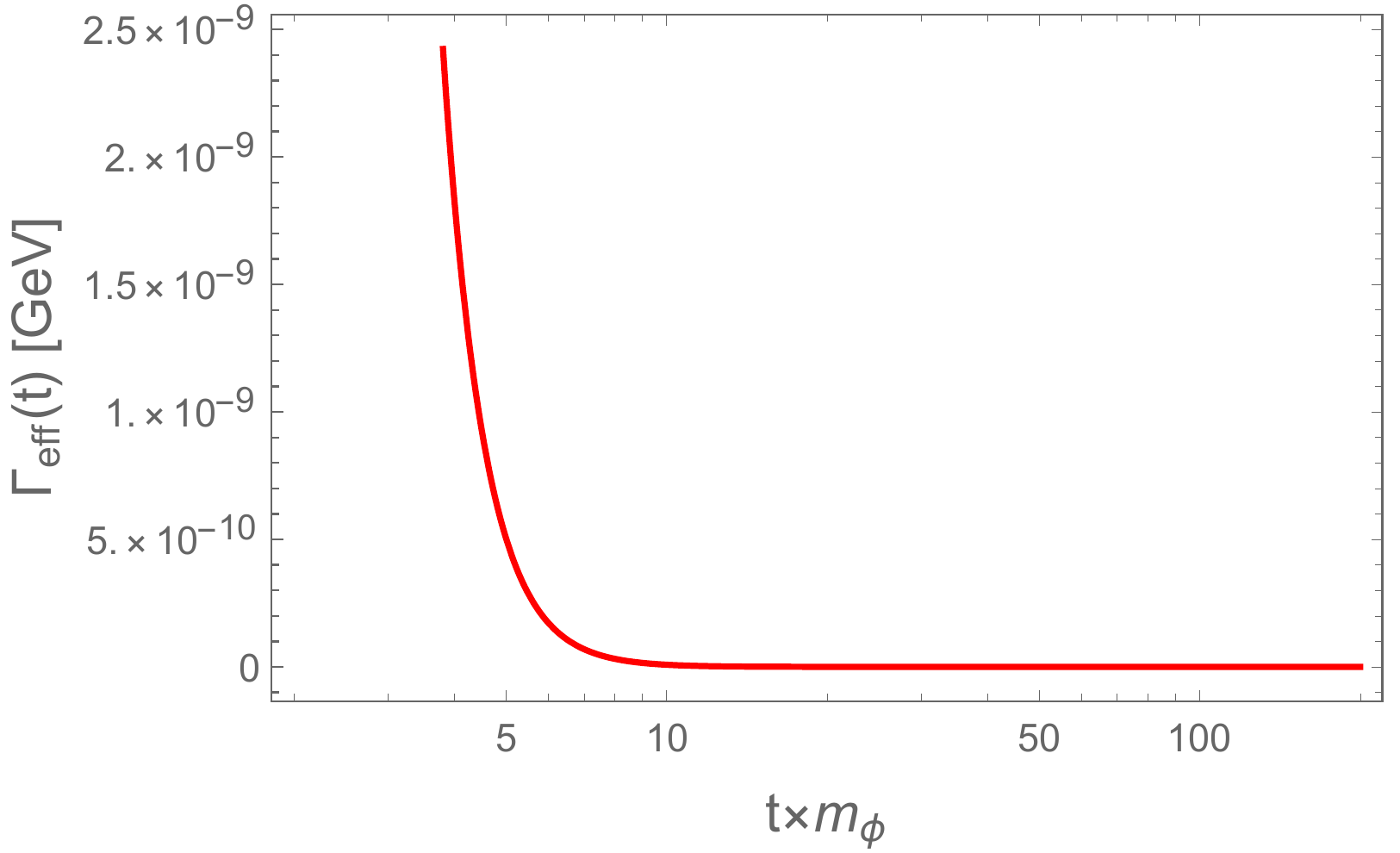}
\caption {The evolution of ALP field, ALP mass, and ALP decay width. The parameter set are: $\theta_0=0.95$, $n_2=5$, $c_r\approx0.1356$, and $f=1.0476\times10^{15} \rm{GeV}$.}
\label{fig:osc}
\end{center}
\end{figure*}

From the left of Fig.~\ref{fig:osc}, we find that the amplitude of the oscillation field decreases with $t\times m_\phi$ rapidly.
With the help of Eq.~($\ref{Eq:fieldosc}$), the effective mass $m_{eff}$ is shown in the middle of Fig.~\ref{fig:osc}, which implies, in the reheating epoch, the effective mass of the field will undergo a steep decline evolution. As decay width $\sim m_{eff}^3$, the reduction effect is further reflected in the effective decay width, which is shown in the right of Fig.~\ref{fig:osc}. We can find that the decay width is large at the beginning of the field oscillation and then rapidly drops to a very low level, which provides the possibility for the inflaton to reheat the universe and then decouple as a long-lived DM.

After determining the effect decay width, we can use Boltzmann equations to study the evolution of energy density during reheating, as follows:
\begin{align}\label{Boltzmann}
&\dot{\rho}_\phi + 3H\rho_\phi = -\Gamma_{eff}\rho_\phi, \nonumber \\
&\dot{\rho}_r+4H\rho_r = \Gamma_{eff}\rho_\phi,
\end{align}
where $\rho_\phi$ and $\rho_r$ denote the energy density of the inflaton and radiation, respectively.
As the ALP potential is dominated by the quadratic term near the potential minimum, the ALP energy density $\rho_\phi$ may redshifts as matter, thus the coefficient of the first line of Eq.~(\ref{Boltzmann}) should be $3$.

\begin{figure}[t]
\begin{center}
\includegraphics[scale=0.48]{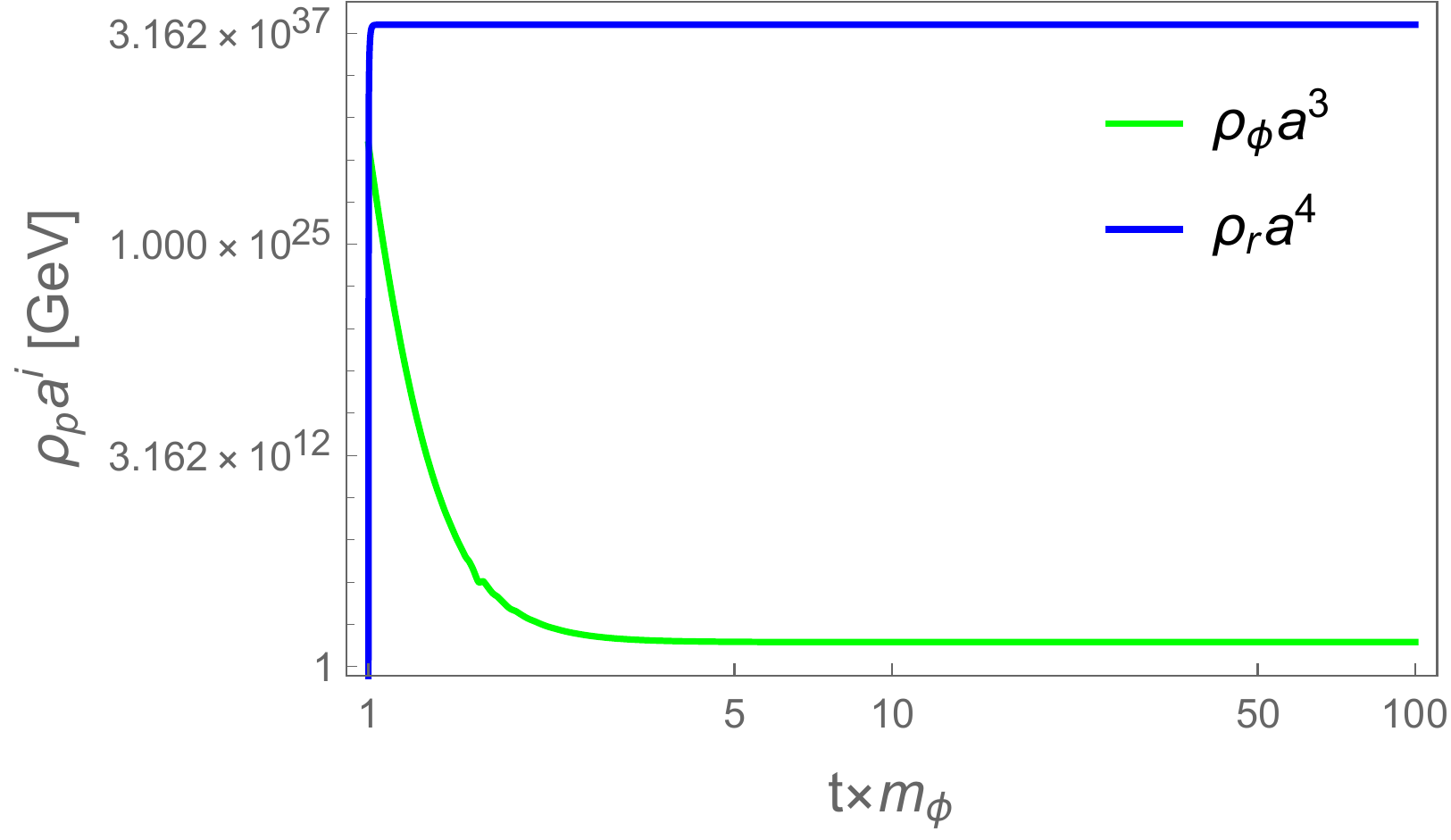}
\caption {The evolution of the comoving energy densities for the inflaton and radiation with the same parameter set as Fig.~\ref{fig:osc}.}
\label{fig:enden}
\end{center}
\end{figure}

With the previous settings, we can solve the Boltzmann equations to get the energy evolution information, which is shown in Fig.~\ref{fig:enden}.
In which we find that almost all the inflaton energy is immediately converted to the radiation due to the decay rate $\Gamma_{eff}$ is much larger than $H$ at the beginning of oscillation, which means instantaneous reheating takes place, and then both the inflaton and the radiation energy density with correspond scale factor remain stable due to the decay rate $\Gamma_{eff}$ goes down fast and then close to $0$, which can be clearly seen from the right of Fig.~\ref{fig:osc}.
Thus ALP does achieve instantaneous heating the universe and may decouple as a DM.

\begin{figure}[t]
\begin{center}
\includegraphics[scale=0.31]{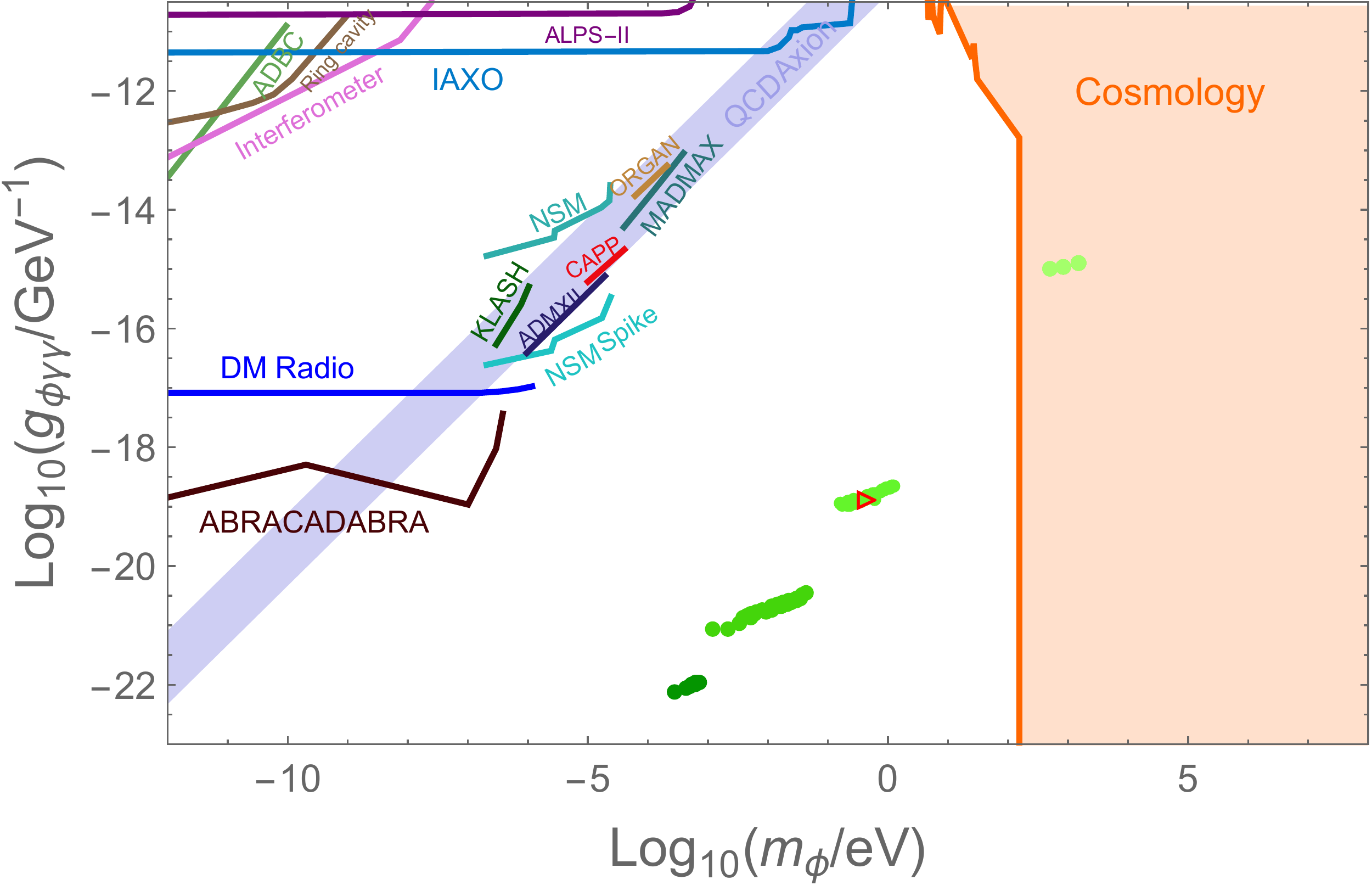}
\caption {The $m_\phi-g_{\phi \gamma\gamma}$ figure for $n_2=3, 5, 7, 9$. In which the darker the green color, the bigger the $n_2$. All the green points are bounded from the BBN temperature for the $T_{RH}$ and DM relic density $\Omega h^2=0.12 $. Those colored lines are the projected experimental reach~\cite{Blinov:2019rhb}. The red triangle is the illustration point for ALP reheating and DM. Note that the case of $n_2=1$ is excluded entirely.}
\label{fig:mg1}
\end{center}
\end{figure}
In the standard cosmology, after the reheating is over, the universe will expand adiabatically, then the ratio of the ALP number density to entropy density $n_\phi/s$ is conserved. Therefore, the ALP DM relic density today can be expressed as~\cite{Blinov:2019rhb}:
\begin{align}
\Omega_{DM} =\frac{\rho_{\phi}(t_{RH})}{\rho_c}\bigg(\frac{T_0}{T_{RH}}\bigg)^3\frac{g_{*S}(T_0)}{g_{*S}(T_{RH})},
\end{align}
where $\rho_c \approx 10^{-5}\,h^2 \mathrm{GeV/cm}^3$ with $h\approx 0.68$~\cite{Tanabashi:2018oca}, $T_0\approx2.7 \rm{K}$. The reheating temperature $T_{RH}$ is defined as the temperature that of the universe when $H(t) \approx \Gamma(t)$~\cite{Chung:1998rq}.
Consider $\rho_{r}=({\pi^2}/{30})g(T)T^4$, the reheating temperature in this case can be estimated as $T_{RH}\sim 10^{2} $$~\rm{GeV}$, which satisfies the reheating temperature lower bound from Big Bang Nucleosynthesis (BBN) $T_{BBN}\sim 10^{-2}~\rm{GeV}$~\cite{Steigman:2007xt}.

After further considering the bound from BBN temperature and DM relic density $\Omega_{DM} h^2=0.12$~\cite{Ade:2015xua}, we plot the feasible points with $n_2= 3, 5, 7, 9$ by using the green points in Fig.~\ref{fig:mg1}. In which the darker green color, the bigger the $n_2$. We find that there's no point here in case $n_2=1$ due to the lager mass of ALP will decay completely in the reheating phase and can not serve as the DM. The case $n_2=3$ is ruled out completely by the cosmology constraints that include Optical, EBL, $\chi_{\rm{ion}}$, X-Rays and DM limits, more details can be obtained in the Ref.~\cite{Cadamuro:2012rm}. The case $n_2=5, 7, 9$ survive for the GNI model.

\section{Summary}\label{Summary}

In this paper, we propose a GNI model with the parameters $n_1$, $n_2$, $\epsilon$, and $C$. We find that the parameter $n_1$ can only be equal to $0,2$ due to the constraints of roll inflation and true vacuum condition for potential, which bring the conditions of the maximal and minimal value for the potential.
When the parameters ($n_1,n_2$) take specific values, GNI can be simplified to some other models.
Specifically, the GNI model can reduce to NI with $(n_1=1,n_2=1)$, and other generalization of natural inflation model with $(n_1=1)$.
Besides, the GNI model with $n_1=2$ will take on unique characters. It not only contain the functions of other generalization of natural inflation model, but also produce both large- and small-inflation, so we study it in depth.

In that case, we find that the ALP mass is insensitive to the decay constant $f$ and parameter $\epsilon$, but sensitive to $n_2$,
because the mass is proportional to $(-\epsilon)^{(n_2-1)/2}$ with $0<\epsilon<1$.
By changing $n_2$ from $1$ to $9$, we can get the mass of ALP from Sub-eV to Super-GeV under the CMB data limits.
ALP with a small mass is a long-lived particle that can serve as a DM.

When inflation is finished, the ALP field will oscillate damped near the lowest potential, and the amplitude of the ALP field will decrease rapidly.
Therefore, in terms of $\Gamma \sim m_{eff}^3(\phi_{amp})$, most of the ALP energy can be converted into radiation energy instantaneously to complete the heating of the universe and leaving part energy as the long-lived DM particle. After taking into account the CMB and other cosmological constraints, The GNI model can address ALP inflation and DM for $n_2=5, 7, 9$.

\section*{Acknowledgment}
We thank Luca Visinelli for helpful discussions on ALP inflation and dark matter.
This work was supported in part by the China Postdoctoral Science Foundation under Grant No.~2019TQ0329 and No.~2020M670476, the National Key R\&D Program of China No.~2017YFA0402204, and the National Natural Science Foundation of China (NSFC) No.~11825506, No.~11821505, No.~U1738209, No.~11851303 and No.~11947302. LB is supported by the National Natural Science Foundation of China under grant No.11605016 and No.11947406, and the Fundamental Research Funds for the Central Universities of China (No. 2019CDXYWL0029).

\bibliographystyle{arxivref}

\end{document}